\documentclass{osa-article}
\journal{ol}
\articletype{preprint}

\newcommand{\be}[0]{\begin{equation}}
\newcommand{\ee}[0]{\end{equation}}
\newcommand{\bea}[0]{\setlength\arraycolsep{2pt}\begin{eqnarray}}
\newcommand{\eea}[0]{\end{eqnarray}}
\newcommand{\ui}{\mathrm{i}}

\newcommand{\br}{\mathbf{r}}

\newcommand{\bk}{\mathbf{k}}

\usepackage{lipsum}  

\begin{document}

\title{Miniature 120-beam coherent combiner with 3D printed optics for multicore fiber based endoscopy}

\author{
	Siddharth Sivankutty, \authormark{1,*,\textdagger}
	Andrea Bertoncini,\authormark{2,*}
	Victor Tsvirkun,\authormark{1},
	Naveen Gajendra Kumar, \authormark{3}
	Gaelle Brévalle,\authormark{4}
	Géraud Bouwmans,\authormark{4}
	Esben Ravn Andresen, \authormark{4}
	Carlo Liberale,\authormark{2,5,**}
	Hervé Rigneault \authormark{1,**}
}

\address{
  \authormark{1} Aix Marseille Univ, CNRS, Centrale Marseille, Institut Fresnel, Marseille, France \\ 
  \authormark{2} Biological and Environmental Science and Engineering, King Abdullah University of Science and Technology (KAUST), Saudi Arabia \\
  \authormark{3}Aix Marseille Univ, CNRS, Centrale Marseille, Institut Fresnel, Turing Centre for Living systems, Marseille, France \\
  \authormark{4}PhLAM CNRS, IRCICA, Universit\'{e} Lille 1, 59658 Villeneuve d’Ascq Cedex, France\\
  \authormark{5}Computer, Electrical and Mathematical Sciences and Engineering Division, King Abdullah University of Science and Technology (KAUST), Saudi Arabia\\
  \authormark{*} Both authors contributed equally to this work \\
  \authormark{\textdagger} Now at PhLAM CNRS, IRCICA.\\
  \authormark{**} Corresponding authors: carlo.liberale@kaust.edu.sa, herve.rigneault@fresnel.fr
}

\begin{abstract*}
We report high efficiency, miniaturized, ultra-fast coherent beam combining with  3D printed micro-optics directly on the tip of a multicore fiber bundle. The highly compact device foot-print (180 micron diameter) facilitates its incorporation into a minimally invasive ultra-thin nonlinear endoscope to perform two-photon imaging. 
\end{abstract*} \\

Imaging at a cellular resolution deep inside the body is a holy grail for medical and research efforts over the world. Scattering in tissue is by far the most detrimental feature that precludes the optical imaging of organs at high resolution \cite{kubby2019wavefront}. In two-photon scanning microscopy \cite{denk1997photon}, the use of longer wavelengths mitigates the tissue scattering and generates a confined focal volume. This results in excellent lateral and axial resolution even at depths of about a 100 $\mu$m. The advent of adaptive optics and wavefront shaping have pushed the imaging routinely down to depths of hundreds of microns in tissues \cite{kubby2019wavefront}. 

On the other hand, medical endoscopes routinely image organs at cellular resolution at much greater depths. However, their applicability is limited to imaging hollow organs given their bulky size ($\geq 1$mm). Hence, there is great interest along two fronts - the deployment of nonlinear imaging techniques in endoscopes and a drastic miniaturization in the size of the  endoscopes to be as minimally invasive as possible \cite{lombardini_high-resolution_2018}. In the recent years, fiber-based lensless endoscopes have emerged, wherein concepts from adaptive optics have enabled the reduction of the endoscope right down to the size of a bare optical fiber with no moving parts at the distal end of fiber. While remarkable demonstrations of lensless endoscopes with multimode fibers and linear imaging contrasts have been reported \cite{turtaev2018high,papadopoulos_focusing_2012,ohayon2018minimally}, non-linear imaging methods are scarce \cite{andresen2013two, sivankutty_ultra-thin_2016, kim_adaptive_2016, conkey_lensless_2016}. 

In our earlier works \cite{andresen_ultrathin_2016}, we have demonstrated that multi-core fiber bundles (MCF) are  ideal candidates in the context of multi-photon microscopy to deliver flexible nonlinear imaging devices. In contrast to traditional fiber bundles, MCFs feature homogeneous single-mode cores which are widely spaced on the bundle. This allows for several advantages such as extremely low cross-talk \cite{andresen2013toward}, very low inter-core group delay dispersion \cite{andresen_measurement_2015} allowing for the delivery for ultrashort pulses and to perform fast and artifact-free two-photon imaging limited only by diffraction \cite{sivankutty_extended_2016}. In addition, they also allow for fast scanning and non-interferometric calibration \cite{sivankutty_single-shot_2018} making them well suited for deployment in a non-laboratory setting. We have demonstrated a key further step towards the realization of flexible lensless endoscopes with a helically twisted MCF \cite{tsvirkun_flexible_2019} which also relies upon sparse and widely spaced cores. For a detailed overview, we refer the readers to \cite{andresen_ultrathin_2016}. 

\begin{figure}
  \includegraphics[width=\linewidth]{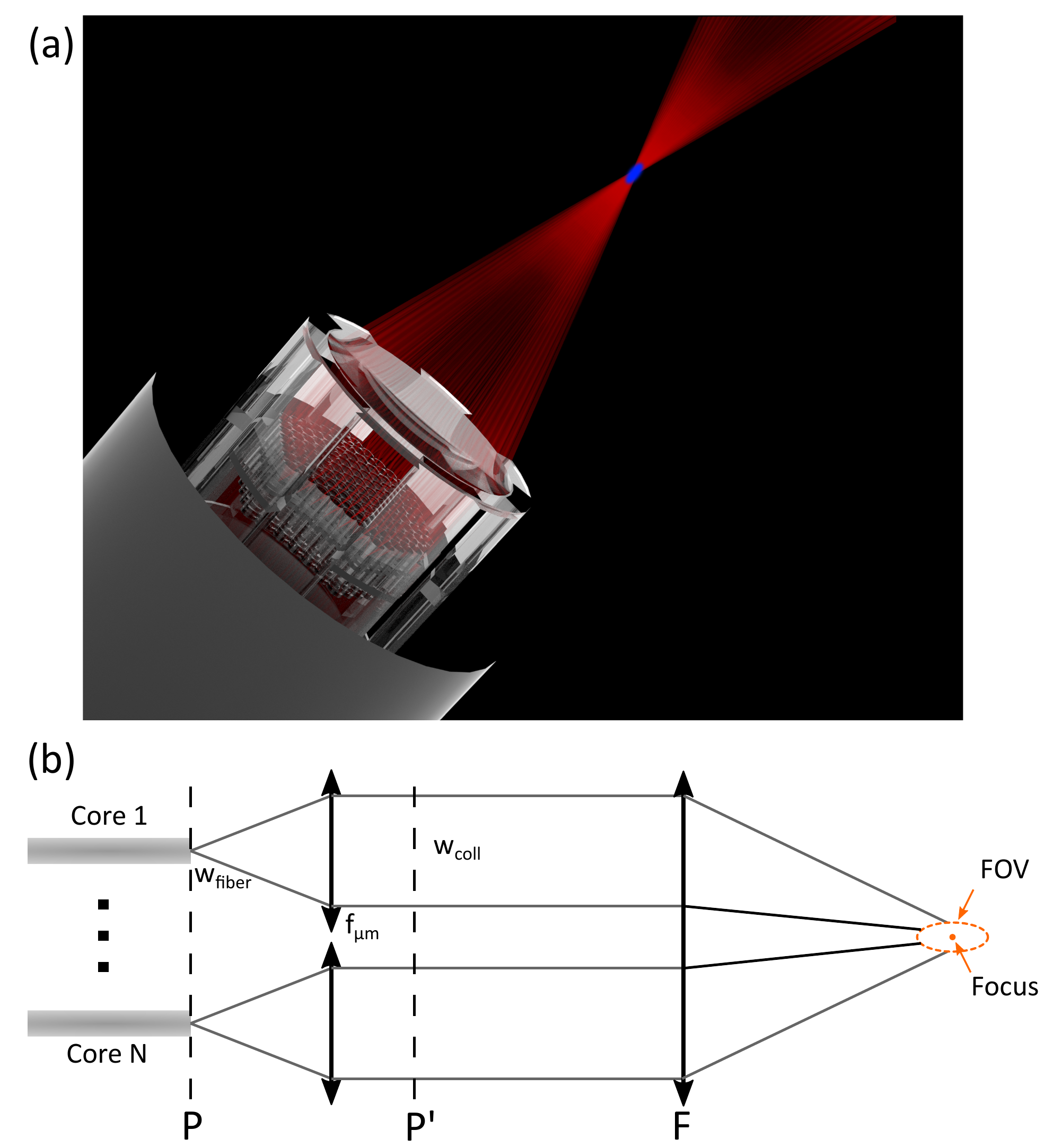}
  \caption{a) A 3D rendering of the micrometric hyper-telescope imprinted on the tip of a fiber b) a conceptual schematic of the same.}
  \label{fig:Concept}
  \end{figure}

Despite these advantages of MCF over conventional multimode fibers (MMF), the sparse spacing of the cores results in one major disadvantage - the relative power $\eta$ in the central focal spot is extremely low ($\approx 0.01$). In this letter, we describe this issue in the context of lensless endoscopes and report the design and fabrication of a highly miniaturized 3D printed optical system at the tip of a multi-core fiber (Fig.~\ref{fig:Concept}a). This results in a gain of x35 in the $\eta$ efficiency,thereby addressing a long-standing roadblock to power scaling and high SNR imaging in ultrathin nonlinear endoscopes.

Two-photon imaging requires a highly intense spatio-temporal focal spot for the quasi-simultaneous absorption of two photons by the fluorophore to generate a fluorescence signal \cite{helmchen2005deep}. Scanning the focus across the sample and detecting the signal on a single pixel detector results in the formation of an image. However in contrast to linearly-excited fluorescence, two-photon absorption requires higher excitation photon flux. In the context of MCF based lensless endoscopes, the generation of such a focal spot is the result of co-phasing the individual beams arising out of each of the cores \cite{andresen2013two}. This results in a constructive interference of the beamlets resulting in a focused spot in the far-field. Let us consider a MCF with $N$ single mode core cores at co-ordinates $\br_n$ for $n=1,...,N$, each emitting a Gaussian beam  with complex amplitude $A$. The  field at the end-face of the fiber can be written as \cite{sivankutty_extended_2016}
\begin{align}
  \mathrm{E(r)} = \mathrm{A(r)} \otimes \sum_{n=1}^{N} \delta (r- \br_n)
\label{eqn:AF}
\end{align}
and the resulting far-field intensity pattern can be expressed in terms of the Fourier transform of Eq.~\ref{eqn:AF} as 

\begin{align}
I(k) = |A(k) \sum_{n=1}^{N} \exp\left(\ui \bk\cdot \br_n \right)|^2
\label{eqn:Far}
\end{align}
In other words, the far-field intensity pattern $I(k)$ is proportional to the 2D power spectrum of the complex field $E(r)$ and the focus is simply the DC term or the electric field strength evaluated at $k=0$, the zero-order spatial frequency.
However, even in the case of perfectly co-phased beamlets, it is not possible to direct all the energy into the central spot. This arises from the fact that the composite near-field $E(r)$ at the MCF endface is modulated in amplitude by presence of the discrete cores and is zero elsewhere. 

As a consequence of this, non-zero Fourier components (higher spatial frequencies) appear in the far-field intensity pattern $I(k)$.
Thus, it is impossible to direct all the energy exiting the cores into a singular diffraction limited focal spot. This is the underlying issue for the poor $\eta$ in our earlier demonstrations. This is evident in the appearance of satellite peaks when the cores are arranged in a periodic grid as in \cite{andresen2013two}. We note that an aperiodic layout of the cores as in \cite{sivankutty_extended_2016} improves the contrast by diminishing the highly intense satellite peaks by redistributing the energy to other non-zero orders, but it does not improve the efficiency of signal generation in the focus. Indeed to enhance the energy delivery into the focus, in addition to co-phasing, the amplitude of the near-field $E(r)$ should resemble a continuous field without the sharp demarcations due to the discrete and well separated cores.
We note a few alternative strategies and their implications here:   i) increasing the laser power would quickly lead to nonlinear effects such as self-phase modulation and temporal broadening of the pulse which counteract the signal generation; ii) increasing the number of cores would lead to a drastic increase in the size of the device; iii)  bringing the cores closer to each other 
entails several disadvantages \cite{conkey_lensless_2016,andresen_ultrathin_2016}, such as cross-talk and mode-coupling which is detrimental to bending insensitivity, flexibility, field of view, fast scanning and calibration.

In this context, we seek to increase $\eta$ by synthetically improving the fill-factor of a sparse pupil (fiber bundle in this case). This has been examined earlier in the context of astronomy as the hyper-telescope \cite{labeyrie1996resolved} for combining light from several telescope arrays and more recently in coherent laser beam combining as tiled-aperture filling \cite{brignon2013coherent,ramirez_coherent_2015}. The key idea is to collimate the beamlets out of each fiber core with a microlens array , thereby increasing its mode-field diameter. Hence, the synthetic pupil at plane P'  has an improved fill-factor given by the ratio $\dfrac{w_{coll}}{w_{fiber}}$, where $w_{fiber}$ and $w_{coll}$ are the beam waist at the core output and after the corresponding microlens, respectively (Fig.~\ref{fig:Concept}b). These expanded and collimated beamlets are then refocused down with a single lens F to generate a far-more intense focus as compared to that obtained with an unstructured MCF (Fig.~\ref{fig:Concept}b). 

In this report, we demonstrate the design and fabrication of a hyper-telescope at the tip of a custom designed 180$\mu$m diameter aperiodic Fermat spiral MCF detailed in \cite{sivankutty_nonlinear_2018} (Fig.~\ref{fig:Concept}a). The relevant parameters of the MCF for design consideration here are the 3.6 $\mu$m mode-field diameter and the aperiodic layout of the cores with an average distance of 11.8 $\mu$m. We modeled the performance of the system with a non-paraxial wave optics model and taking into account factors such as the vignetting by the apertures of the microlens array \cite{brignon2013coherent}.
In addition to the combining efficiency, we optimized the focal spot size and the total area in which the spot can be steered by implementing phase-tilts with the SLM. 
The optimal focal length ($f_{\mu m}$) of the micro-lenses in the array is found to be 39 $\mu$m, the focusing lens focal length (F) to be 800 $\mu$m and the distance between the two to be 100 $\mu$m. In this framework, we estimate for the ideal case an $\eta$ of 0.5, with a focal spot  size of 4.8 $\mu$m  and a field of view of about 60 $\mu$m. This appears as a good trade-off between working distance, resolution and field of view. 

\begin{figure}
\centering
\includegraphics[width=\columnwidth]{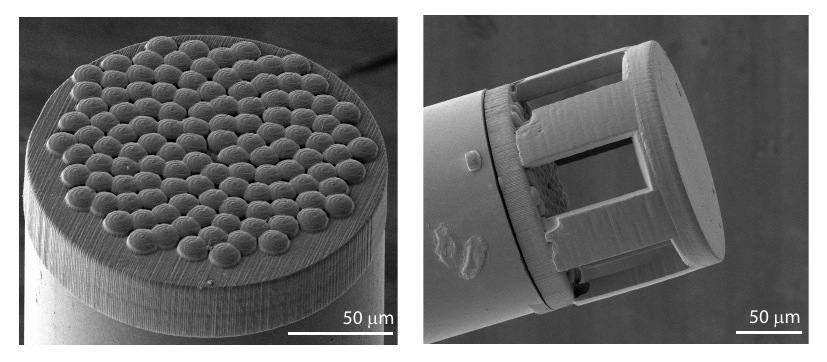}
\caption{Scanning electron micrographs of a) the microlens array only and b) the entire hyper-telescope where the top focusing lens has been printed 100 $\mu$m over the microlens array.}
\label{fig:SEM}
\end{figure}

\begin{figure}[ht]
\includegraphics[width=\columnwidth]{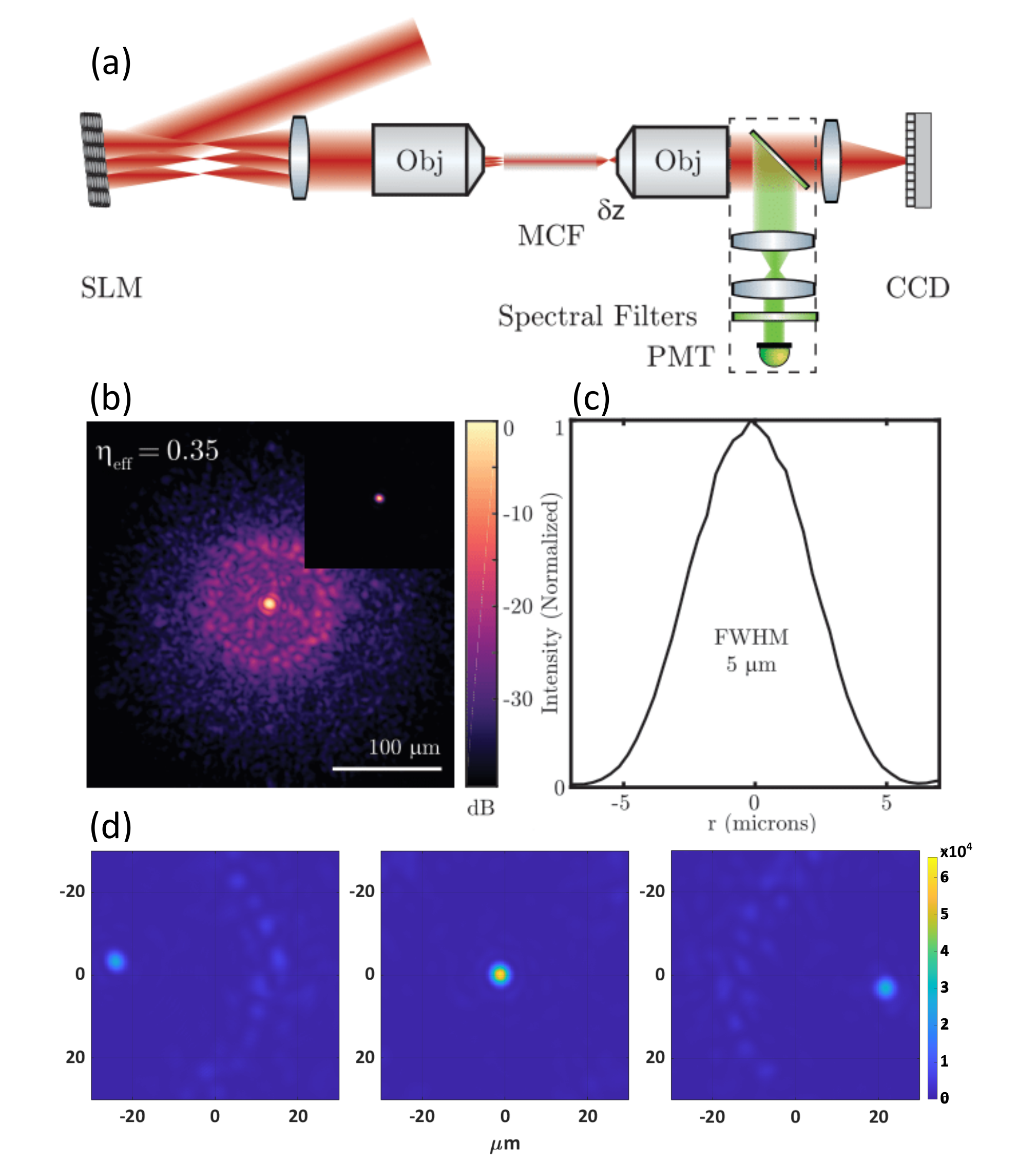}
\caption{a) A simplified schematic used for characterization and imaging; b) a log-scale image of the focal spot (inset - Linear); c) the linear point-spread function (d) focal spot point-spread function at left edge, center and right edge of the field of view, from left to right respectively} 
\label{fig:Setup}
\end{figure}

Given the small diameter of the MCF, the pitch and the scale of the structures, we use two-photon lithography as a tool to directly fabricate the designed hyper-telescope on the tip of the MCF. This is a powerful tool demonstrated for in-situ fabrication of a variety of  micro-optics on fibers in the context of imaging, inter-connects and photonic-circuits \cite{gissibl2016two,thiele20173d,bertoncini20203d}.
In particular, we use a commercial 3D printer (Photonic Professional GT, Nanoscribe), with the proprietary IP-S photoresist (extinction coefficient 0.1/mm, Nanoscribe) and a 25x 0.8 NA microscope objective (Zeiss).
The system is equipped with a camera that allows to align the 3D printed structure with the MCF. The alignment of the lenslet array on the aperiodic layout of cores is facilitated by a custom-developed computer vision algorithm that detects the cores positions and prints one lenslet at a time on top of each core.
In our first designs, the arrayed lenslets and the focusing top lens F are designed as simple spherical and plano-convex lenses, respectively.
The radius of curvature (ROC) of the lenslets is 14 $\mu$m and the ROC of the top lens is 400 $\mu$m. 
Usually, the 3D printing coordinates are defined by slicing 2D planes along the printing direction. However, this strategy can result in a staircase-like surface or lead to overexposure \cite{park2005subregional} in areas that are almost flat, like the central part of lenses. To avoid these problems, we defined the 3D printed coordinates for the lenses as a single layer of concentric circles with a constant pitch in diameter, placed on different planes to form a spherical lens. 

The optical performance of the 3D printed hyper-telescope on the custom MCF was investigated with the setup sketched in Fig.~\ref{fig:Setup}(a) which has been described in detail in \cite{sivankutty_nonlinear_2018}. The key performance indicator reflecting the beam combining efficiency is $\eta$, i.e. the effective power ratio of the central focal spot over the total power. A continuous-wave laser source operating at 1030 nm (2 W, IPG Photonics) impinges upon a spatial light modulator (SLM, X10468 Hamamatsu).  The phase pattern on the SLM is divided into segments with the position of each segment matched to that of a corresponding fibre core. Each of these segments is a sum of: (i) a parabolic term to generate a focus at the respective core entrances, maximising the coupling efficiency, taking care to match the numerical aperture of the cores; (ii) a linear term acting as a blazed diffraction grating such that we work in the first-order to ensure no unmodulated light is coupled into the fiber; (iii) a constant term which controls the phase of the field coupled into the respective cores. After an initial mapping of the SLM to the MCF proximal facet with standard computer vision algorithms, we run an iterative procedure which fine-tunes the position of the center of each segment on the SLM to its corresponding core by maximizing the coupled power. 
The distal end of the fiber and the focal plane of the hyper-telescope combiner is subsequently imaged on to a camera (Flea3, FLIR) with an imaging system (magnification $\approx$ 7.5, numerical aperture = 0.2). A polarizer in front of the camera  transmits only the \textit{s-} polarized light since the polarization state out of the fiber is random \cite{sivankutty_single-shot_2016}. We also ensure that when imaging the focal plane of the combiner, there is no vignetting and all the light is captured on the sensor. 

In order to estimate the combining efficiency and the point spread function, we first measure the relative phase differences between the fiber cores with respect to the central core by phase stepping interferometry.  When these relative phase differences are compensated for by displaying the conjugate phase on the SLM, a focal spot is generated at 800 $\mu$m from the combiner top lens as visualized in Fig. \ref{fig:Setup}(b) with a FWHM of 5 $\mu$m (Fig. \ref{fig:Setup}(c)). 
Since the central focal spot is typically 100 times more intense than the average background speckle, we obtain high-dynamic range images \cite{ohayon2018minimally} such that we carefully account for the power in the background speckle to determine the efficiency $\eta$.

\begin{figure}
  \centering
  \includegraphics[width=\linewidth]{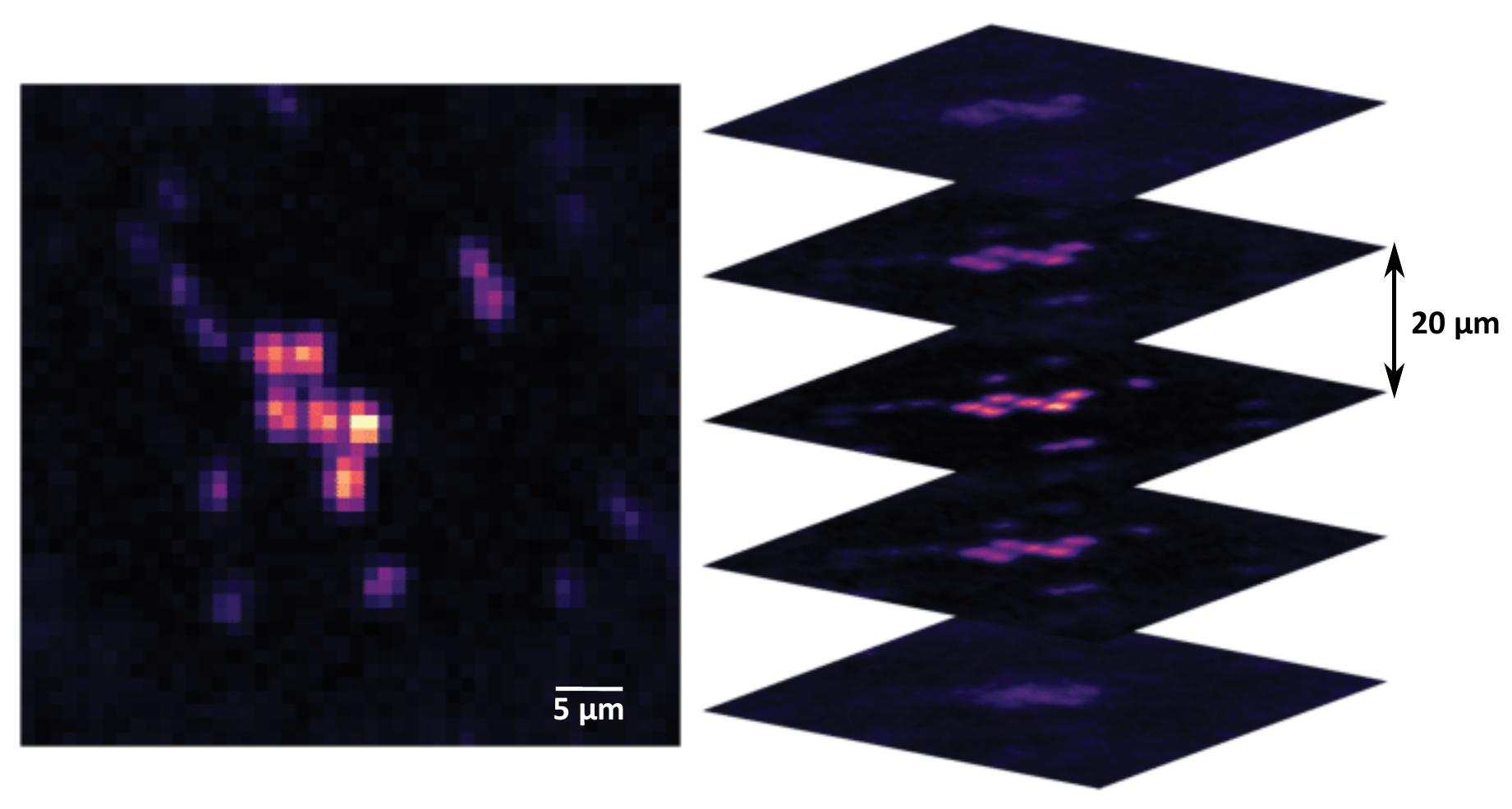}
  \caption{a) In-focus  and b) volume view of two-photon images of 2~$\mu$m microbeads, each of the imaging planes are separated by 20~$\mu$m. Field of view 48 x 48 $\mu$m, integration time per pixel 100 ms (Limited by the SLM update rate)}
  \label{fig:Beads}
  \end{figure}

For our current system, we obtain an experimental efficiency $\eta$ of $\mathrm{0.35}$ and a focal spot size of 5.2 $\mu$m (FWHM). This is in contrast to an efficiency of $\leq 0.01$ in previously reported work with an unstructured MCF with a similar fill-factor \cite{sivankutty_nonlinear_2018,kim_adaptive_2016}. This is 70\% of  the computed efficiency for our given geometry and an observed 8\% broadening of the diffraction limited point-spread function. We primarily attribute this to the top plano-convex lens with 800 $\mu$m focal length  in the hyper-telescope whose field-curvature results in a reduced overlap of the individual fiber modes on the imaging plane. In upcoming iterations, we could account for this in the design of this lens \cite{thiele20173d}. Also, $\eta$ is measured only for the $s$ polarized state due to the fact that the  MCF does not preserve polarization. This is not a key limiting factor, since we could incorporate either polarization maintaining MCFs \cite{kim_adaptive_2016} or implement full polarization control on the SLM. 
Furthermore, we demonstrate that the developed system is entirely compatible for nonlinear imaging. We replace the CW laser source with a 150 femtosecond pulsed laser operating at 920 nm (Chameleon, Coherent Inc) to perform two-photon imaging. We observe no significant pulse broadening due to the micro-optics and the relatively short length of the fiber (200 mm) precludes significant broadening due to inter-core group delay dispersion. Both lateral and axial scanning is performed here  by distal wavefront shaping with the SLM with no mechanical perturbation of the fiber-tip and  the generated fluorescence signal is detected with a PMT (H7421-40, Hammatsu, dotted box in  Fig~ \ref{fig:Setup}(a)). For lateral scanning we phase-tilt only the wavefront coupled in the MCF, taking advantage of the memory effect of our MCF and hyper-telescope combiner. This results in a field of view of 56 $\mu$m defined by the distance at which $\eta$ drops to 50 \% of its max value (Fig. \ref{fig:Setup}(d)). Fig.~\ref{fig:Beads} shows two-photon volume images of 2~$\mu$m beads and attest to the high fidelity imaging of the device. Compared to earlier demonstrations without hyper-telescope that required 100s of milliwatts of total laser power at the distal end of the fiber to get few mW into the focus, we obtained these images with 25 mW of total laser power at the distal end of the fiber that lead to $\sim$9 mW into the focus. This demonstrates a significant enhancement of the sensitivity and opens the route to perform two-photon imaging with lower average powers in the sample volume. The system has a x-y resolution of 5 $\mu$m defined by the focal spot size and a z resolution of 60 $\mu$m. Furthermore, we observed the ability to move the focus over a range of 500~$\mu$m along the z axis without significant changes of the central focal spot size.

We have demonstrated the use of a 3D printed miniaturized hyper-telescope 120-fiber coherent combiner at the tip of a multicore fiber bundle for ultra-thin nonlinear endoscopes. This reduces the footprint of the entire system down to a cylinder of diameter 180 $\mu$m and height 200 $\mu$m with an efficiency of $\eta$=0.35 and a resulting rms error of $\approx \dfrac{\lambda}{25}$. To the best of our knowledge, this is both the largest number of beams combined and the smallest ever device foot-print for coherent beam combiners and its application to imaging systems. This first demonstration indicates the potential of 3D printed micro-optics for scaling up power requirements in applications such as nonlinear imaging and ultra-fast pulsed laser combination.

\section*{Acknowledgements}
We thank Miguel A. Alonso, Institut Fresnel for the stimulating discussions.

\section*{Disclosures}
The authors declare no conflicts of interest.

\section*{Funding Information}

We acknowledge financial support from the Centre National de la Recherche Scientifique (CNRS), Aix-Marseille University (A-M-AAP-ID-17-13-170228-15.22-RIGNEAULT), Turing Centre for Living systems (ANR-16-CONV-0001), ANR grant (ANR-20-CE19-0028-02), NIH R21 EY029406-01, and King Abdullah University of Science and Technology (BAS/1/1064-01-01).

\bibliography{Biblio}
\end{document}